\documentclass[journal]{IEEEtran}
\usepackage{amsmath,bm}

\usepackage{cite}
\ifCLASSINFOpdf
  \usepackage[pdftex]{graphicx}
\else
\fi
\usepackage{amsmath}
\usepackage{amssymb}
\usepackage{array}
\usepackage[caption=false,font=footnotesize]{subfig}

\begin{document}
\title{Burst-Error Propagation Suppression for Decision-Feedback Equalizer in Field-Trial Submarine Fiber-Optic Communications}

\author{Ji Zhou, Chengkun Yang, Dawei Wang, Qi Sui, Haide Wang, Shecheng Gao, Yuanhua Feng, Weiping Liu, Yuelin Yan, Jianping Li, Changyuan Yu and Zhaohui Li% <-this % stops a space

\thanks{This work is supported in part by National Key R\&D Program of China (2018YFB1802300); National Natural Science Foundation of China (62005102, U2001601); Natural Science Foundation of Guangdong Province (2019A1515011059); Fundamental Research Funds for the Central Universities (21619309); Open Fund of IPOC (BUPT) (IPOC2019A001).}

\thanks{J. Zhou, H. Wang, Shecheng Gao, Y. Feng and W. Liu are with Department of Electronic Engineering, College of Information Science and Technology, Jinan University, Guangzhou 510632, China (zhouji@jnu.edu.cn).}% <-this % stops a space

\thanks{C. Yang, D. Wei and Z. Li are with the State Key Laboratory of Optoelectronic Materials and Technologies, School of Electronics and Information Technology, Sun Yat-sen	University, Guangzhou 510275, China.}

\thanks{Y. Yan is with the China Mobile Group Guangdong Co Ltd-Zhuhai Branch, Zhuhai, China.}% <-this % stops a space

\thanks{Q. Sui and Z. Li are with the Southern Marine Science and Engineering Guangdong Laboratory (Zhuhai), Zhuhai, China.}

\thanks{J. Li are with the School of Information Engineering, Guangdong University of Technology, Guangzhou 510006, China.}
\thanks{Changyuan Yu is with the Department of Electronic and Information Engineering, The Hong Kong Polytechnic University, Hong Kong.}
}% <-this % stops a space
\maketitle

\begin{abstract}
In this paper, we present a field-trial C-band 72Gbit/s optical on-off keying (OOK) system over 18.8km dispersion-uncompensated submarine optical cable in the South China Sea. Chromatic dispersion (CD) of 18.8km submarine optical cable causes four spectral nulls on the 36GHz bandwidth of 72Gbit/s OOK signal, which is the main obstacle for achieving an acceptable bit-error-rate (BER) performance. Decision feedback equalizer (DFE) is effective to compensate for the spectral nulls. However, DFE has a serious defect of burst-error propagation when the burst errors emerge due to the unstable submarine environment. Weighted DFE (WDFE) can be used to mitigate the burst-error propagation, but it cannot fully compensate for the spectral nulls because only a part of feedback symbols is directly decided. Fortunately, maximum likelihood sequence estimation (MLSE) can be added after the WDFE to simultaneously eliminate the resisting spectral distortions and implement optimal detection. Compared to the joint DFE and MLSE algorithm, the joint WDFE and MLSE algorithm can effectively suppress the burst-error propagation to obtain a maximum 2.9dB improvement of $\boldsymbol{Q}$ factor and eliminate the phenomenon of BER floor. In conclusion, the joint WDFE and MLSE algorithm can solve the burst-error propagation for the field-trial fiber-optic communications.
\end{abstract}


\IEEEpeerreviewmaketitle

%-------------------------------------------------- Introduction Section -------------------------------------------------------%

\section{Introduction}
With the development of archipelago, the demand of data transmission among islands grows rapidly. Submarine fiber-optic communications are suitable for connecting the islands owing to its high capacity and high reliability. Intensity-modulation and direct-detection (IM/DD) optical systems have advantages of low cost and simple structure \cite{Liu2018, Lange2018, Zhong, cheng2018recent}, which can be easily deployed to effectively connect the nearby islands of archipelago. Compared to standard single-mode fiber (SSMF) in Lab, the submarine optical cable has a higher transmission loss. Therefore, IM/DD optical systems over the submarine optical cables tend to work at C band with relative low transmission losses. However, chromatic dispersion (CD) at C band induces serious spectral distortions, which is the main obstacle to high capacity-distance product for IM/DD optical systems \cite{Chagnon:19,Fu2020,Zhang:18}.

Dispersion-compensation module (DCM) is a common way to accurately compensate for the CD\cite{Eiselt, Zhang:17-1, Eiselt:16}. However, DCM should be customized depending on the length of fiber link. Meanwhile, high-gain optical amplifier is required to compensate for the losses of fiber link and DCM. The complex signal modulations have been proposed to replace one-dimensional IM schemes for eliminating CD, but which require two electrical links including two digital-to-analog converters (DACs), two electrical drivers and dual-drive Mach-Zehnder modulator (DD-MZM), or in-phase/quadrature (I/Q) MZM\cite{Zhang, Zhang:17, Li}. Recently, Tomlinson-Harashima precoding has been proposed to resist CD-caused spectral distortions, which requires precoding and decoding modules at both the transmitter and receiver ends, respectively \cite{Rath, Hu, Xin}. In our previous work, we proposed adaptive channel-matched detection (ACMD) including polynomial nonlinear equalizer (PNLE), decision-feedback equalizer (DFE) and maximum likelihood sequence estimation (MLSE) for 64Gbit/s on-off keying (OOK) signal over 100km SSMF in the Lab, which can flexibly and effectively compensate for CD-caused spectral distortions for IM/DD optical systems \cite{wang2020adaptive}.

In this paper, we present a field trial of C-band 72Gbit/s optical OOK system over 18.8km dispersion-uncompensated submarine optical cable in the South China Sea. CD of 18.8km submarine optical cable causes four nulls on the 36GHz spectrum of 72Gbit/s OOK signal. DFE can effectively compensate for the spectral nulls due to the poles of its frequency-domain transfer function\cite{tang2020digital, tang2020low, Zhou:21}. However, the burst-error propagation is the major drawback of DFE, which limits its applications in practical fiber-optic communications. For mitigating the burst-error propagation, we propose the weighted DFE (WDFE) cooperating with MLSE. Compared to the classical DFE, the WDFE can shorten the length of burst errors, which is conducive to the sequence detection of MLSE. Meanwhile, MLSE can compensate for the resisting spectral distortions when the WDFE cannot fully compensate for the spectral nulls. The main contributions of this paper are as follows:
\begin{itemize}
	\item The joint WDFE and MLSE algorithm is proposed to effectively eliminate the serious CD-caused spectral nulls and suppress the burst-error propagation. 
	\item Field-trial 72Gbit/s OOK system is demonstrated over 18.8km dispersion-uncompensated submarine optical cable in the South China Sea. Compared to the joint DFE and MLSE algorithm, the joint WDFE and MLSE algorithm can effectively suppress the burst-error propagation to obtain maximum 2.9dB improvement of $Q$ factor.
\end{itemize}

The remainder of this paper is organized as follows. In Section \ref{Principle of WDFE}, the principles of the joint WDFE and MLSE algorithm are analysed in detail. In Sections \ref{Section3}, the field-trial submarine fiber-optic communications are demonstrated to verify the performance of error propagation suppression using the joint WDFE and MLSE algorithm. Finally, the paper is concluded in Section \ref{Conclusions}.

\section{Joint WDFE and MLSE Algorithm} \label{Principle of WDFE}
\begin{figure}[!t]
	\centering
	\includegraphics[width=0.8\linewidth]{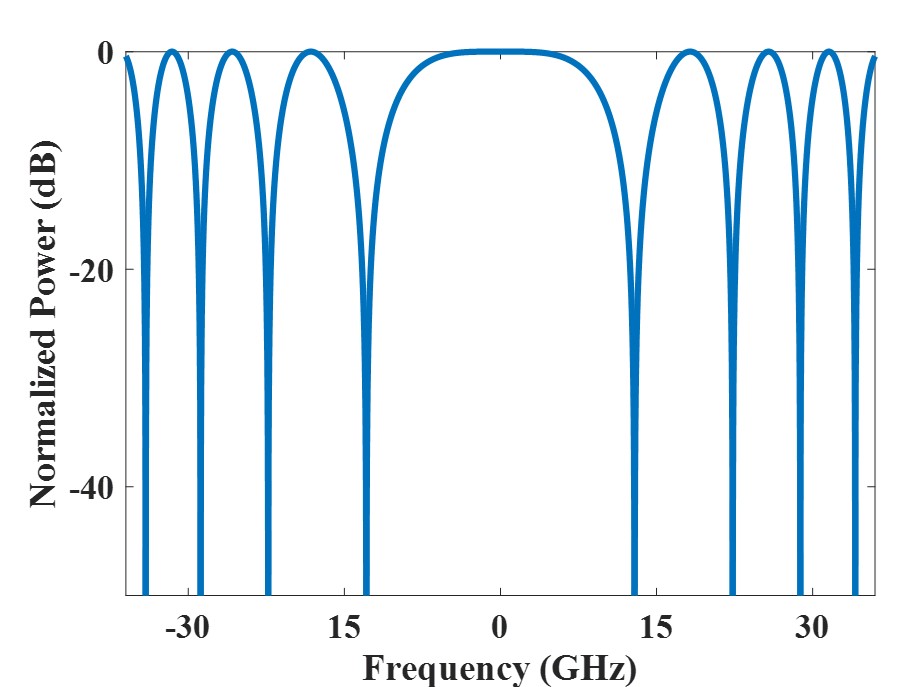}
	\caption{Theory frequency-domain response of 18.8-km dispersive channel.}
	\label{FiberResponse}
\end{figure}
\begin{figure}[!t]
	\centering
	\includegraphics[width=0.98\linewidth]{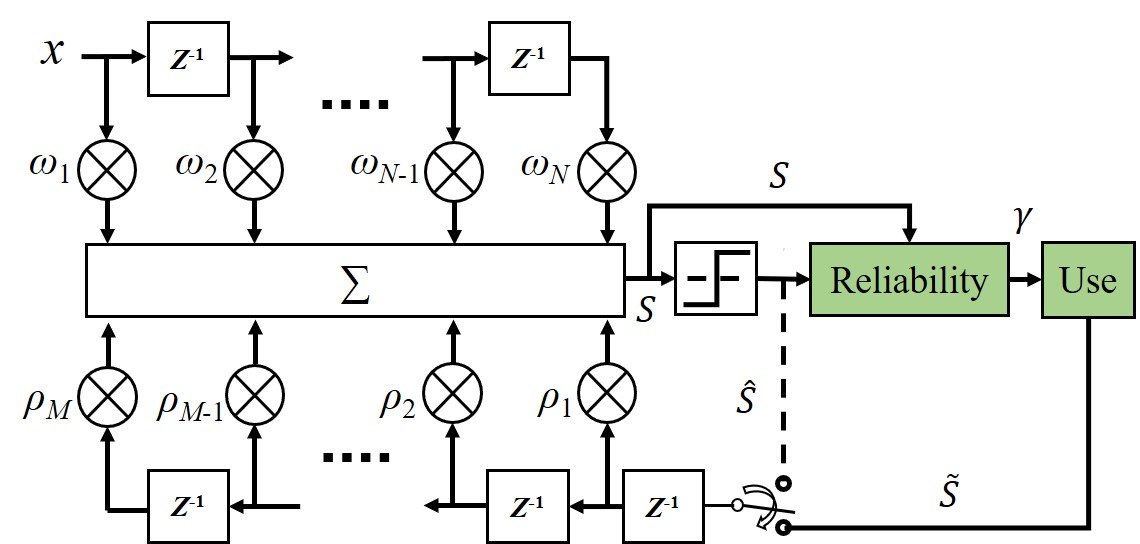}
	\caption{The block diagram of DFE and WDFE. When the tap is on the top (bottom) of switch, DFE (WDFE) is implemented.}
	\label{WDFE}
\end{figure}

In this section, the principle of the joint DFE and MLSE algorithm is demonstrated to compensate the CD-caused spectral nulls. The frequency-domain response of dispersive channel can be expressed as \cite{zhou2016transmission}
\begin{equation}
H(f) = \cos(2\pi^2\beta_2Lf^2)
\end{equation}
where $\beta_2$ is the group velocity dispersion and $L$ is the fiber length. Specially, the theory frequency-domain response of 18.8km dispersive channel is shown in Fig. \ref{FiberResponse}. There are four spectral nulls within the frequency range from 0 to 36 GHz. Furthermore, $H(f)$ of dispersive channel can be decomposed into
\begin{equation}
\begin{aligned}
H(f) = \sum_{n=0}^{\infty}(-1)^{n}\frac{(2\pi^2 \beta_2 L f^2)^{2n}}{(2n)!}= \sum_{n=0}^{\infty}a(n)f^{4n}.
\end{aligned}
\end{equation}
Therefore, $H(f)$ can be expressed as summation polynomials with the frequency-domain null points. For compensating the CD-caused distortions, the frequency-domain transfer function of equalizer should be reciprocal of summation polynomials. As an auto-regressive filter, the frequency-domain transfer function of DFE is the reciprocal of summation polynomials \cite{ingle2016digital}, which can be defined as
\begin{equation}
\begin{aligned}
H(f) &= \frac{1}{ \prod_{n=0}^{\infty}[1-p(n)e^{-j2\pi f}]}
=\frac{1}{\sum_{k=0}^{\infty}b(k)f^{k}}.
\end{aligned}
\end{equation}
In the ideal conditions, DFE completely compensates for the CD-caused spectral distortions when the $\bm{b}$ in $H(f)$ of DFE is converged to match the $\bm{a}$ in $H(f)$ of dispersive channel by an adaptive algorithm. Unfortunately, DFE suffers from burst-error propagation due to the decision-feedback error symbols. The burst-error propagation seriously degrades the performance of sequence detection including MLSE and forward error correction (FEC) \cite{song2003reduced, xie2009improving}, which leads to a phenomenon of bit-error-rate (BER) floor. WDFE can mitigate burst-error propagation by introducing reliability \cite{palicot2000weighted, palicot2008performance}. The block diagram of WDFE is shown in Fig. \ref{WDFE}. It works as the classical DFE when the switch of feedback turns up and it serves as the WDFE when the switch turns down. Therefore, the feedback symbols of the classical DFE are the decision outputs $\hat{s}$. The feedback signal of the WDFE is a combination of output $s$ and the decision output $\hat{s}$.

\begin{figure}[!t]
	\centering
	\includegraphics[width=0.9\linewidth]{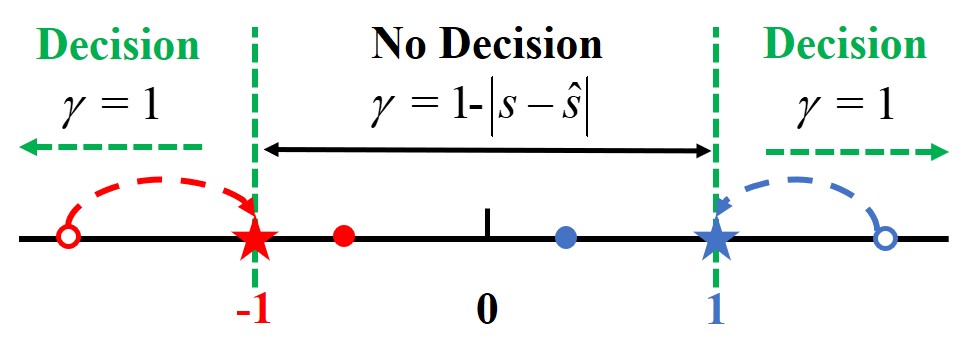}
	\caption{The calculation of reliability value $\gamma$ for the PAM2 constellation.}
	\label{RELIABILITY}
\end{figure}

The major difference between the classical DFE and WDFE is the addition of two new blocks: $\it{Reliability}$ and $\it{Use}$. The $\it{Reliability}$ block calculates a reliability value $\gamma$ of $s$, which is similar with a belief or a likelihood measurement. As shown in Fig. \ref{RELIABILITY} for two-level pulse-amplitude modulation (PAM2), when the symbol (solid circle) is between -1 and 0 or between 0 and 1, the distance between the symbol and its hard decision is firstly calculated by
\begin{equation}
\Delta d = |s - \hat{s}|.
\label{eq4}
\end{equation}
Then the reliability value $\gamma$ is calculated by
\begin{equation}
\gamma = 1 - \Delta d.
\label{eq5}
\end{equation}
When the symbol gets to the border (i.e., 0), its reliability value is closer to $0$. When the symbol is near the constellation point (i.e., $\pm 1$), its reliability value is around $1$. Additionally, when the symbol (hollow circle) is less than $-1$ or greater than $1$, the error probability is small and the symbol can be directly decided. Therefore, the reliability value can be set to $1$. Therefore, Eq. (\ref{eq5}) can be revised as
\begin{equation}
\gamma=\left\{\begin{array}{cc}
1-\left|s-\hat{s}\right|, & \left|s\right| \leqslant 1 \\
1, &\left|s\right|>1
\end{array}\right..
\end{equation}
 
The $\it{Use}$ block uses the reliability value $\gamma$ to obtain the feedback symbol. The feedback symbol is a combination of the output $s$ and the decision output $\hat{s}$, which can be defined as
\begin{equation}
\tilde{s}=f\left(\gamma\right) \hat{s}+\left[1-f\left(\gamma\right)\right] s
\label{eq6}
\end{equation}
where $f\left(\cdot\right)$ is a monotonically increasing threshold function. The simplest $f\left(\cdot\right)$ is the identity function. Specially, Eq. (\ref{eq6}) can be calculated by 
\begin{equation}
\tilde{s}=\gamma \hat{s}+\left(1-\gamma\right) s.
\end{equation}
The WDFE can be considered as a compromise proposal between infinite impulse response (IIR) filter and classical DFE. When the $\gamma$ is 1, WDFE can be considered as classical DFE. When the $\gamma$ is 0, WDFE can be considered as IIR. When the $\gamma$ is small, the error propagation is prevented due to the small weight of the decision symbol. When the $\gamma$ is large, the weight of the decision symbol is increased with a high degree of confidence. The tap coefficient vectors $\bm{\omega}$ and $\bm{\rho}$ can be updated by decision-directed least mean square algorithm,
\begin{equation}
\bm{\omega}=\bm{\omega}+ \mu_1 \gamma e\bm{x}, 
\end{equation}
\begin{equation}
\bm{\rho}= \bm{\rho} + \mu_2 \gamma e \tilde{\bm{s}}
\end{equation}
where $\mu_1$ and $\mu_2$ are step sizes and error $e$ is calculated by
\begin{equation}
e=\hat{s}-s. 
\end{equation}

\begin{figure}[t]
	\centering
	\includegraphics[width = 0.9\linewidth]{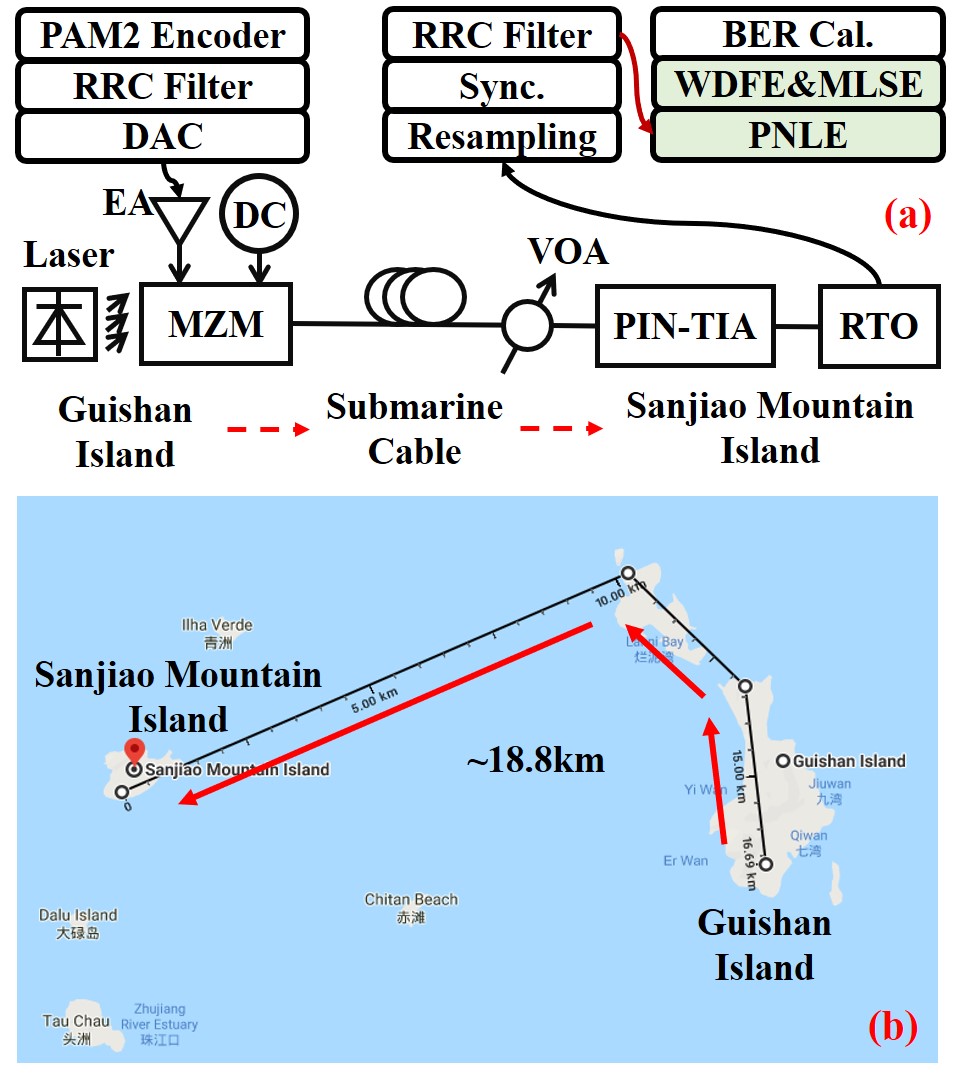}
	\caption{(a) Experimental setups for field-trial submarine fiber-optic communications; (b) The path of commercial submarine optical cable between Guishan island and Sanjiao mountain island in South China Sea.}
	\label{EX}
\end{figure}

In the WDFE, only a part of feedback symbols is directly decided. Therefore, the WDFE cannot fully compensate for the spectral nulls. The resisting spectral distortions still degrade the BER performance. Following the WDFE, MLSE is employed to compensate for the resisting ISI and implement optimal detection. In order to detect the most likely transmitted symbols in MLSE, the minimum Euclidean distance $D(\boldsymbol{s}, \boldsymbol{t})$ should be obtained, which can be expressed as
\begin{equation}
D(\boldsymbol{s}, \boldsymbol{t}) =\sum_{k}[s_k-\sum_{i = 0}^{P} \omega_{i} \cdot t_{k-i}]^{2}.
\end{equation}
where $\boldsymbol{s}$ is the output of WDFE. $\boldsymbol{t}$ is the desired transmitted symbol. $\boldsymbol{\omega}$ is the tap coefficient of ISI. $P$ is the memory length of MLSE.

\section{Field-Trial Submarine Fiber-Optic Communications} \label{Section3}
In this section, a field-trial submarine fiber-optic communications are demonstrated to verify the performance of the joint WDFE and MLSE algorithm for suppressing the burst-error propagation.

\subsection{Experimental Setups}
Figure \ref{EX}(a) shows the experimental setups for the field-trial submarine fiber-optic communications. Firstly, at the transmitter end, the digital PAM2 frame was generated by offline processing. The PAM2 frame was filtered by the a root-raised-cosine (RRC) filter with the 0.02 roll-off factor to implement Nyquist pulse shaping. Then the digital frame was uploaded into DAC with the sampling rate of 90GSa/s and 3dB bandwidth of 16 GHz. After the resampling operation, the link rate of the generated PAM2 signal was 72Gbit/s and the net rate was approximately 63Gbit/s (72Gbit/s$\times$77240/82240/(1+7\%) $\approx$ 63Gbit/s) when the length of training symbols was set to 5000 and hard-decision FEC (HD-FEC) with 7\% overhead was used. After being amplified by an electrical amplifier (EA) (Centellax OA4SMM4), a 40Gbps MZM @ single drive mode (Fujitsu FTM7937EZ) was used to modulate the amplified PAM2 signal on optical carrier at 1550.116nm for generating optical OOK signal. 2V direct-current (DC) bias was applied on the MZM.

Then the optical OOK signal was fed into the submarine optical cable. The launch optical power was set to 6.25dBm. The link loss was approximately 0.49 dB/km, which is much larger than the 0.2dB/km loss of SSMF in Lab. Fig. \ref{EX}(b) shows the path of commercial submarine optical cable between Guishan island and Sanjiao mountain island in South China Sea. The total length of submarine optical cable is about 18.8km. At the receiver end, a variable optical attenuator (VOA) was used to adjust the received optical power (ROP). The optical OOK signal was converted into an electrical signal by a 31GHz PIN with trans-impedance amplifier (PIN-TIA) (Finisar MPRV1331A). The electrical signal was fed into a 80GSa/s real-time oscilloscope (RTO) with cutoff bandwidth of 36GHz to implement analog-to-digital conversion. The cutoff bandwidth of RTO limits the maximum rate of OOK signal, which was set to 72Gbit/s. The digital OOK signal was recovered by offline processing, including resampling, synchronization, RRC matching filter, 3-order PNLE, WDFE, post filter (PF), MLSE, and BER calculation.

\begin{figure}[!t]
	\centering
	\includegraphics[width=\linewidth]{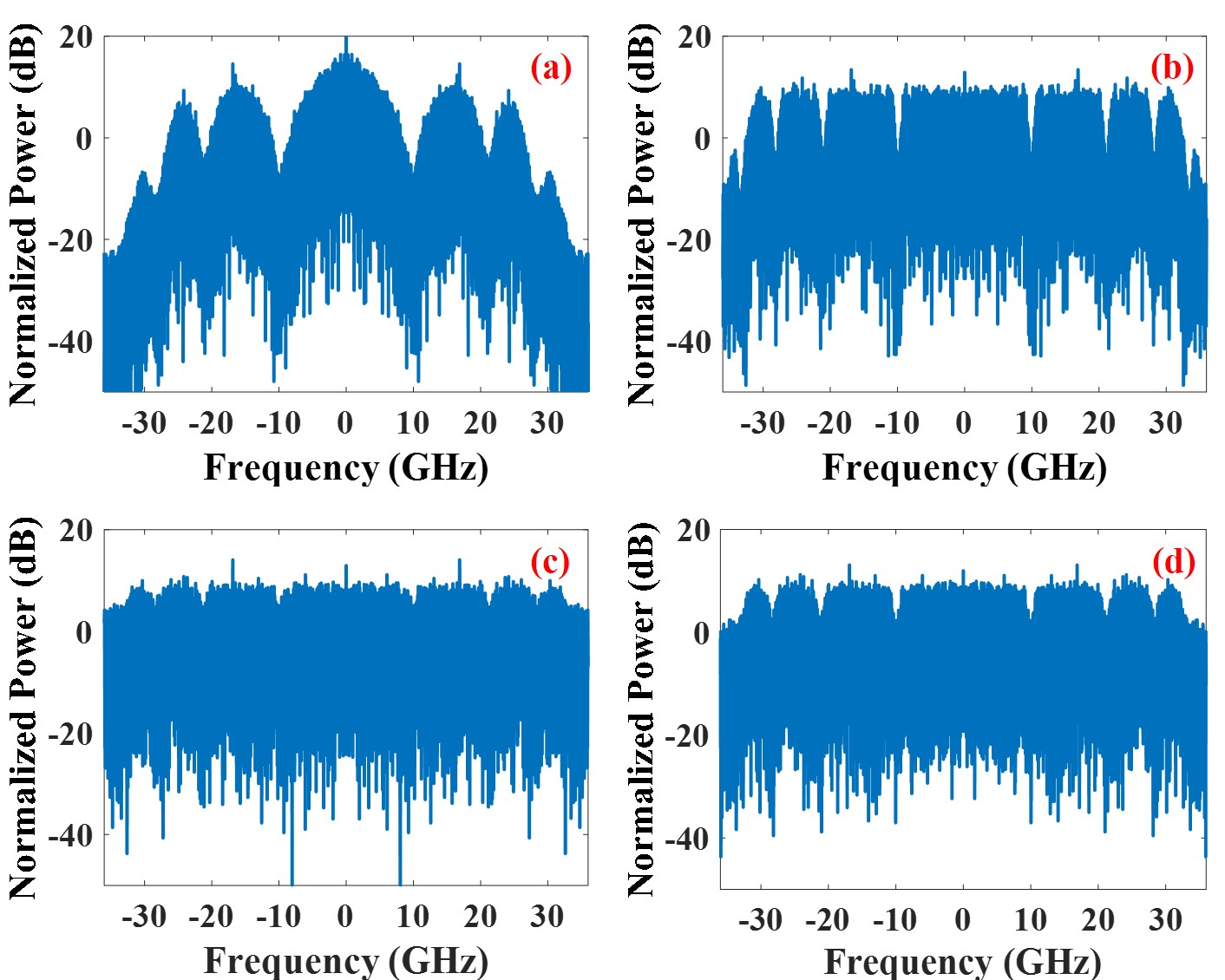}
	\caption{Signal spectra of (a) the received 72Gbit/s OOK signal, (b) the output of PNLE, (c) the output of classical DFE and (d) the output of WDFE.}
	\label{SPECTRA}
\end{figure}

\subsection{Results and Discussions}
Figure \ref{SPECTRA} depicts the signal spectra of (a) the received 72Gbit/s OOK signal, (b) the output of the PNLE, (c) the output of the classical DFE and (d) the output of the WDFE. The spectrum of the received signal indicates that the signal has suffered from serious distortions as shown in Fig. \ref{SPECTRA}(a). After the 18.8km dispersion-uncompensated submarine optical cable, the received signal suffers from four spectral nulls caused by the CD, which is consistent with theoretical analysis in Section \ref{Principle of WDFE}. Meanwhile, the high-frequency part of the signal spectrum fades fast due to the limited bandwidth of devices, especially the RTO. In the 3-order PNLE, linear, square and cubic terms compensate for the high-frequency linear distortions, the signal-signal beat interference and the 3-order nonlinear distortions, respectively \cite{zhou2019adaptive}. The tap number of PNLE is set to (81, 71, 51). As Fig. \ref{SPECTRA}(b) shows, PNLE can compensate for the main spectral distortions except for a part of the high-frequency distortions. The frequency-domain transfer function of PNLE has no pole. Therefore, PNLE cannot compensate for the CD-induced spectral nulls. Fortunately, DFE has the frequency-domain transfer function with poles, which can effectively eliminate the spectral nulls and the residual high-frequency distortions. The tap numbers of DFE and WDFE are both set to (71, 51). As Fig. \ref{SPECTRA}(c) depicts, after the classical DFE, the CD-induced spectral nulls can be effectively compensated. However, as Fig. \ref{SPECTRA}(d) shows, WDFE only compensates for a part of the spectral nulls, which agrees well with the theory analysis in Section \ref{Principle of WDFE}. 

\begin{figure}[!t]
	\centering
	\includegraphics[width=0.9\linewidth]{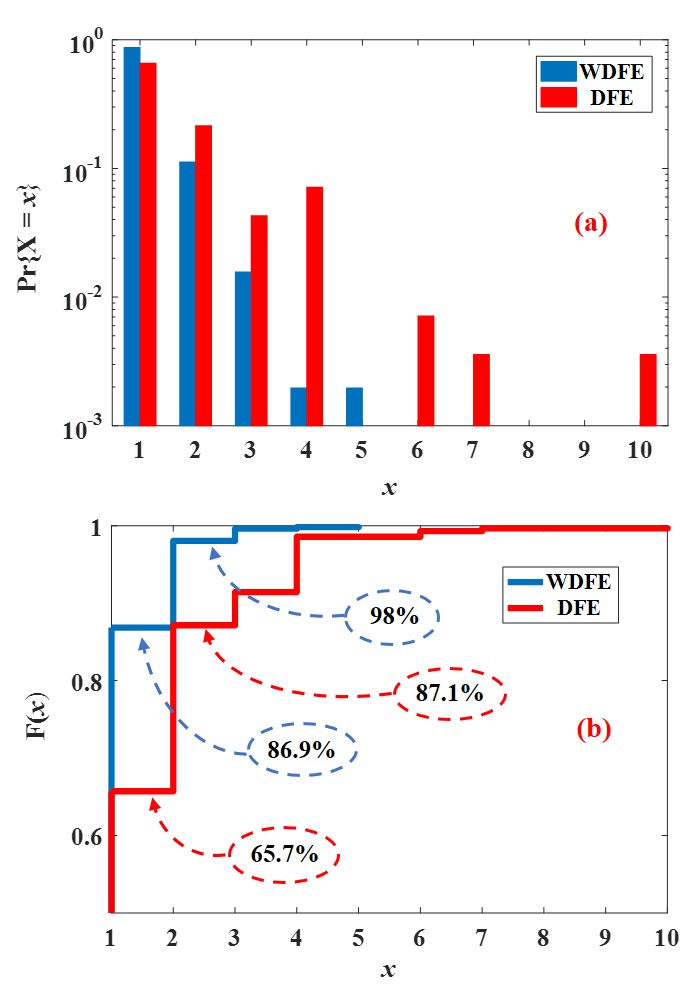}
	\caption{(a) The PDF and (b) the CDF of burst consecutive errors at the ROP of $-4$ dBm using WDFE and the classical DFE, respectively.}
	\label{DF}
\end{figure}

Figure \ref{DF}(a) shows the probability distribution function (PDF) of the burst consecutive errors at the ROP of $-4$ dBm using WDFE or the classical DFE, respectively. The maximum length of burst consecutive errors is 10 when the classical DFE is used, while it can reduced to 5 by using WDFE. Fig. \ref{DF}(b) shows the cumulative distribution function (CDF) of the burst consecutive errors at ROP of $-4$ dBm using WDFE or the classical DFE, respectively. The percentage of single-bit error is 65.7\% using DFE, while it increases to 86.9\% employing WDFE. The total percentage of the single-bit error and two consecutive errors is 98\%, which is $\sim11\%$ more than that using the classical DFE. In conclusion, the WDFE has better performance on the suppression of burst-error propagation.

\begin{figure}[!t]
	\centering
	\includegraphics[width=0.9\linewidth]{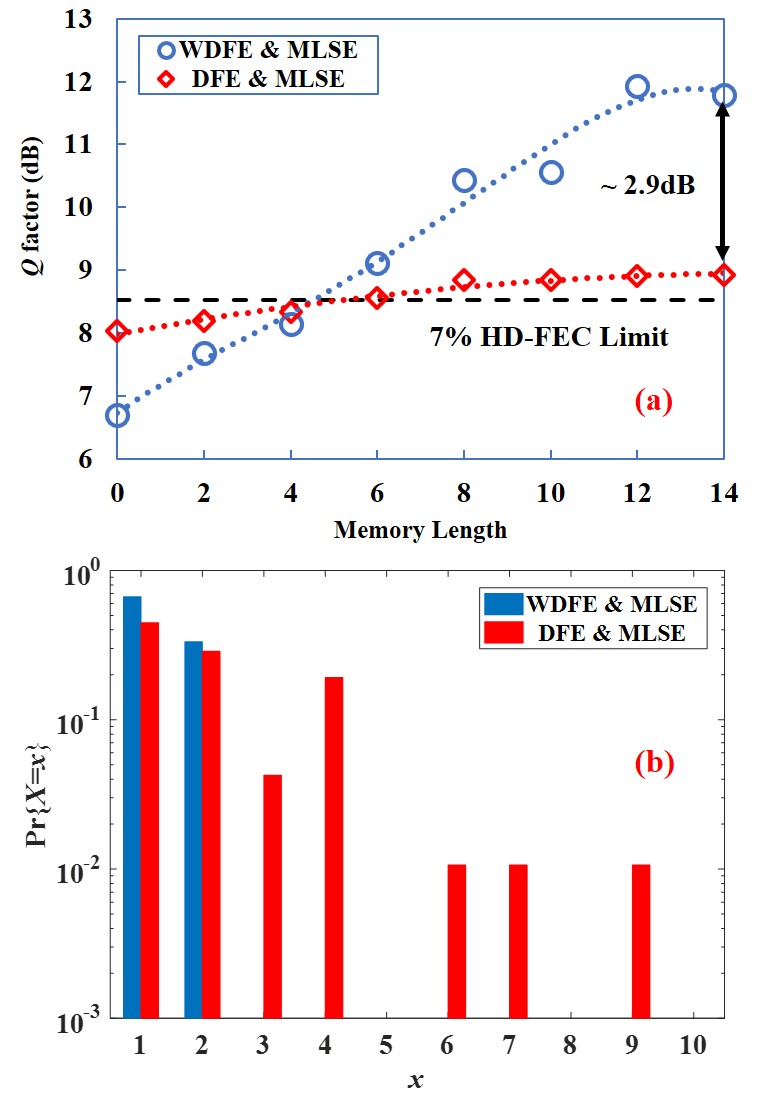}
	\caption{(a) $Q$ factor versus the memory length of MLSE for 72Gbit/s IM/DD optical OOK system over 18.8-km dispersion-uncompensated submarine optical cable when the ROP is set to $-4$ dBm. (b) The PDF of burst consecutive errors at the ROP of $-4$ dBm using joint WDFE and MLSE algorithm and joint DFE and MLSE algorithm, respectively.}
	\label{Q}
\end{figure}

Following the WDFE or classic DFE, MLSE is employed to compensate for the resisting ISI and implement optimal detection. By using the Viterbi searching algorithm, the MLSE obtains the most probable data sequence. The performance of MLSE is sensitive to the length of burst consecutive errors. Fig. \ref{Q} (a) shows the $Q$ factor versus the memory length of MLSE for 72Gbit/s IM/DD optical OOK system over 18.8km dispersion-uncompensated submarine optical cable when the ROP is set to $-4$ dBm. $Q$ factor is calculated from the BER as described in \cite{bergano1993margin}. When the MLSE does not work (i.e., the memory length is equal to 0), the performance of the classic DFE is better than that of WDFE. This is because the classic DFE can compensate more spectral nulls than the WDFE. The $Q$ factor increases with the increase of memory length of MLSE. When the memory length is larger than $6$, the $Q$ factors of two schemes are both higher than the value corresponding to the 7\% HD-FEC limit. Meanwhile, the $Q$ factor of WDFE becomes higher than that of the classic DFE. However, when the memory length of MLSE increases to 14, the $Q$ factor increases less than $1$dB for the case of classic DFE. The main reason is that MLSE cannot deal with large burst consecutive errors due to the serious burst-error propagation of DFE. For the case of WDFE, the $Q$ factor can be improved $\sim5$dB by using MLSE with the memory length of 14, which is $\sim2.9$dB higher than that for the case of the classic DFE. 

Figure \ref{Q}(b) shows the PDF of burst consecutive errors at the ROP of $-4$ dBm using joint WDFE and MLSE algorithm and joint DFE and MLSE algorithm, respectively. After joint DFE and MLSE algorithm, the maximum length of burst consecutive errors is $9$. Therefore, the MLSE with memory length of $14$ cannot obviously eliminate the burst consecutive errors of DFE. However, after joint WDFE and MLSE algorithm, the maximum length of burst consecutive errors is $2$. The MLSE with memory length of $14$ can greatly reduce the burst consecutive errors of WDFE. In conclusion, joint WDFE and MLSE algorithm has better performance on the suppression of burst-error propagation than the joint DFE and MLSE algorithm, achieving to a better BER performance.

\begin{figure}[!t]
	\centering
	\includegraphics[width=0.88\linewidth]{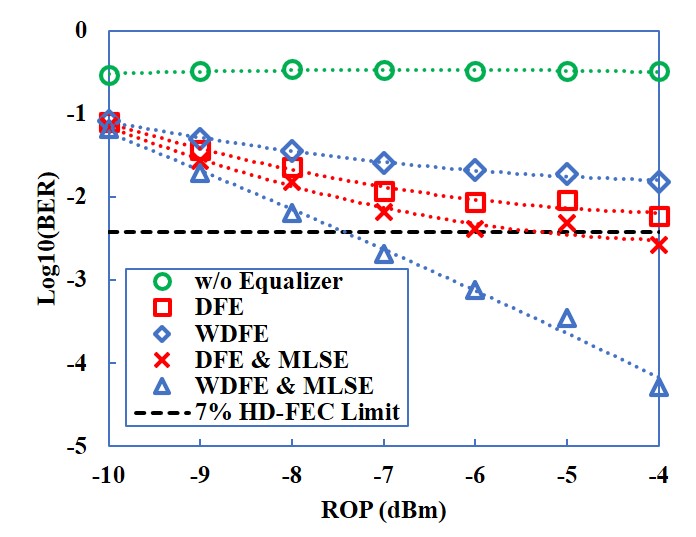}
	\caption{BER versus ROP for 72Gbit/s IM/DD optical OOK system over 18.8-km dispersion-uncompensated submarine optical cable.}
	\label{BER}
\end{figure}

Figure \ref{BER} shows BER versus ROP for 72Gbit/s IM/DD optical OOK system over 18.8km dispersion-uncompensated submarine optical cable. Due to the serious distortions, the BER is near 0.5 when no equalizer is employed. After DFE or WDFE, the BER performance can be improved. The DFE has better performance than WDFE due to its more effective for compensating the spectral nulls. The burst-error propagation in both DFE and WDFE causes the BER floor. The BER cannot achieve the 7$\%$ HD-FEC limit using only DFE or WDFE. The MLSE following the DFE or WDFE can further improve the BER performance. The BER is below the 7$\%$ HD-FEC limit when the joint DFE and MLSE algorithm or the joint WDFE and MLSE algorithm is employed. As the above analysis shows, the MLSE cannot deal with the serious burst-error propagation of DFE. The BER floor still exists when the joint DFE and MLSE algorithm is employed. Owing to the less burst-error propagation, the BER floor can be eliminated and the BER is less than $10^{-4}$ when the joint WDFE and MLSE algorithm is employed. For achieving the BER under 7\% HD-FEC limit, the required ROP using the joint WDFE and MLSE algorithm is $3$dB lower than that using the joint DFE and MLSE algorithm. 

\section{Conclusions}\label{Conclusions}
In this paper, we present a field-trial C-band 72Gbit/s optical OOK system over 18.8km dispersion-uncompensated submarine optical cable in the South China Sea. The DFE is effective to compensate for the CD-caused spectral nulls. However, DFE easily appears a phenomenon of burst-error propagation when the burst errors emerges due to unstable submarine environment. The joint WDFE and MLSE algorithm is proposed to effectively suppress the burst-error propagation. Compared to the joint DFE and MLSE algorithm, the joint WDFE and MLSE algorithm can effectively suppress the burst-error propagation to obtain maximum 2.9dB improvement of $\boldsymbol{Q}$ factor and eliminate the phenomenon of BER floor. For achieving the BER under 7\% HD-FEC limit, the required ROP using the joint WDFE and MLSE algorithm is $3$dB lower than that using joint DFE and MLSE algorithm. In conclusion, the joint WDFE and MLSE algorithm can solve the burst-error propagation for the field-trial fiber-optic communications.

%-------------------------------------------------- Bibliography Section -------------------------------------------------------%

\ifCLASSOPTIONcaptionsoff
  \newpage
\fi

\bibliographystyle{IEEEtran}
\bibliography{sample}

\begin{thebibliography}{10}
\providecommand{\url}[1]{#1}
\csname url@samestyle\endcsname
\providecommand{\newblock}{\relax}
\providecommand{\bibinfo}[2]{#2}
\providecommand{\BIBentrySTDinterwordspacing}{\spaceskip=0pt\relax}
\providecommand{\BIBentryALTinterwordstretchfactor}{4}
\providecommand{\BIBentryALTinterwordspacing}{\spaceskip=\fontdimen2\font plus
\BIBentryALTinterwordstretchfactor\fontdimen3\font minus
  \fontdimen4\font\relax}
\providecommand{\BIBforeignlanguage}[2]{{%
\expandafter\ifx\csname l@#1\endcsname\relax
\typeout{** WARNING: IEEEtran.bst: No hyphenation pattern has been}%
\typeout{** loaded for the language `#1'. Using the pattern for}%
\typeout{** the default language instead.}%
\else
\language=\csname l@#1\endcsname
\fi
#2}}
\providecommand{\BIBdecl}{\relax}
\BIBdecl

\bibitem{Liu2018}
G.~N. {Liu}, L.~{Zhang}, T.~{Zuo}, and Q.~{Zhang}, ``{IM/DD Transmission
  Techniques for Emerging 5G Fronthaul, DCI, and Metro Applications},''
  \emph{Journal of Lightwave Technology}, vol.~36, no.~2, pp. 560--567, 2018.

\bibitem{Lange2018}
S.~Lange, S.~Wolf, J.~Lutz, L.~Altenhain, R.~Schmid, R.~Kaiser, M.~Schell,
  C.~Koos, and S.~Randel, ``{100 GBd Intensity Modulation and Direct Detection
  with an InP-based Monolithic DFB Laser Mach-Zehnder Modulator},''
  \emph{Journal of Lightwave Technology}, vol.~36, no.~1, pp. 97--102, 2018.

\bibitem{Zhong}
K.~{Zhong}, X.~{Zhou}, J.~{Huo}, C.~{Yu}, C.~{Lu}, and A.~P.~T. {Lau},
  ``{Digital Signal Processing for Short-Reach Optical Communications: A Review
  of Current Technologies and Future Trends},'' \emph{Journal of Lightwave
  Technology}, vol.~36, no.~2, pp. 377--400, 2018.

\bibitem{cheng2018recent}
Q.~Cheng, M.~Bahadori, M.~Glick, S.~Rumley, and K.~Bergman, ``Recent advances
  in optical technologies for data centers: a review,'' \emph{Optica}, vol.~5,
  no.~11, pp. 1354--1370, 2018.

\bibitem{Chagnon:19}
M.~Chagnon, ``{Direct-detection Technologies for Intra-and Inter-data Center
  Optical Links},'' in \emph{Optical Fiber Communication Conference}.\hskip 1em
  plus 0.5em minus 0.4em\relax Optical Society of America, 2019, p. W1F.4.

\bibitem{Fu2020}
Y.~{Fu}, D.~{Kong}, H.~{Xin}, S.~{Jia}, K.~{Zhang}, M.~{Bi}, W.~{Hu}, and
  H.~{Hu}, ``{Piecewise Linear Equalizer for DML Based PAM-4 Signal
  Transmission Over a Dispersion Uncompensated Link},'' \emph{Journal of
  Lightwave Technology}, vol.~38, no.~3, pp. 654--660, 2020.

\bibitem{Zhang:18}
K.~Zhang, Q.~Zhuge, H.~Xin, W.~Hu, and D.~V. Plant, ``{Performance comparison
  of DML, EML and MZM in dispersion-unmanaged short reach transmissions with
  digital signal processing},'' \emph{Optics Express}, vol.~26, no.~26, pp.
  34\,288--34\,304, Dec 2018.

\bibitem{Eiselt}
N.~{Eiselt}, J.~{Wei}, H.~{Griesser}, A.~{Dochhan}, M.~H. {Eiselt},
  J.~{Elbers}, J.~J.~V. {Olmos}, and I.~T. {Monroy}, ``{Evaluation of Real-Time
  8$\times$56.25 Gb/s (400G) PAM-4 for Inter-Data Center Application Over 80 km
  of SSMF at 1550 nm},'' \emph{Journal of Lightwave Technology}, vol.~35,
  no.~4, pp. 955--962, 2017.

\bibitem{Zhang:17-1}
J.~Zhang, J.~Yu, and H.-C. Chien, ``{EML-based IM/DD 400G
  (4$\times$112.5-Gbit/s) PAM-4 over 80 km SSMF Based on Linear
  Pre-Equalization and Nonlinear LUT Pre-Distortion for Inter-DCI
  Applications},'' in \emph{Optical Fiber Communication Conference}.\hskip 1em
  plus 0.5em minus 0.4em\relax Optical Society of America, 2017, p. W4I.4.

\bibitem{Eiselt:16}
N.~Eiselt, J.~Wei, H.~Griesser, A.~Dochhan, M.~Eiselt, J.-P. Elbers, J.~J.~V.
  Olmos, and I.~T. Monroy, ``{First Real-Time 400G PAM-4 Demonstration for
  Inter-Data Center Transmission over 100 km of SSMF at 1550 nm},'' in
  \emph{Optical Fiber Communication Conference}.\hskip 1em plus 0.5em minus
  0.4em\relax Optical Society of America, 2016, p. W1K.5.

\bibitem{Zhang}
Q.~Zhang, N.~Stojanovic, C.~Xie, C.~Prodaniuc, and P.~Laskowski,
  ``{Transmission of single lane 128 Gbit/s PAM-4 signals over an 80 km SSMF
  link, enabled by DDMZM aided dispersion pre-compensation},'' \emph{Optics
  Express}, vol.~24, no.~21, pp. 24\,580--24\,591, Oct 2016.

\bibitem{Zhang:17}
Q.~Zhang, N.~Stojanovic, T.~Zuo, L.~Zhang, C.~Prodaniuc, F.~Karinou, C.~Xie,
  and E.~Zhou, ``{Single-Lane 180 Gb/s SSB-Duobinary-PAM-4 Signal Transmission
  over 13 km SSMF},'' in \emph{Optical Fiber Communication Conference}.\hskip
  1em plus 0.5em minus 0.4em\relax Optical Society of America, 2017, p. Tu2D.2.

\bibitem{Li}
Z.~{Li}, M.~S. {Erkilinc}, K.~{Shi}, E.~{Sillekens}, L.~{Galdino}, B.~C.
  {Thomsen}, P.~{Bayvel}, and R.~I. {Killey}, ``{168 Gb/s/$\lambda$
  Direct-Detection 64-QAM SSB Nyquist-SCM Transmission over 80 km Uncompensated
  SSMF at 4.54 b/s/Hz net ISD using a Kramers-Kronig Receiver},'' in
  \emph{European Conference on Optical Communication}, 2017, p. Tu.2.E.1.

\bibitem{Rath}
R.~{Rath}, D.~{Clausen}, S.~{Ohlendorf}, S.~{Pachnicke}, and W.~{Rosenkranz},
  ``{Tomlinson-Harashima Precoding For Dispersion Uncompensated PAM-4
  Transmission With Direct-Detection},'' \emph{Journal of Lightwave
  Technology}, vol.~35, no.~18, pp. 3909--3917, 2017.

\bibitem{Hu}
Q.~{Hu}, K.~{Schuh}, M.~{Chagnon}, F.~{Buchali}, S.~T. {Le}, and H.~{Bulow},
  ``{50 Gb/s PAM-4 Transmission Over 80-km SSMF Without Dispersion
  Compensation},'' in \emph{{European Conference on Optical Communication}},
  2018, p. We3F.6.

\bibitem{Xin}
H.~Xin, K.~Zhang, D.~Kong, Q.~Zhuge, Y.~Fu, S.~Jia, W.~Hu, and H.~Hu,
  ``{Nonlinear Tomlinson-Harashima precoding for direct-detected double
  sideband PAM-4 transmission without dispersion compensation},'' \emph{Optics
  Express}, vol.~27, no.~14, pp. 19\,156--19\,167, Jul 2019.

\bibitem{wang2020adaptive}
H.~Wang, J.~Zhou, D.~Guo, Y.-H. Feng, C.~Yu, W.~Liu, and Z.~Li, ``{Adaptive
  Channel-Matched Detection for {C}-Band 64-{G}bit/s Optical {OOK} System over
  100-km Dispersion-Uncompensated Link},'' \emph{Journal of Lightwave
  Technology}, 2020.

\bibitem{tang2020digital}
X.~{Tang}, Y.~{Qiao}, Y.~W. {Chen}, Y.~{Lu}, and G.~K. {Chang}, ``{Digital Pre-
  and Post-Equalization for C-Band 112-Gb/s PAM4 Short-Reach Transport
  Systems},'' \emph{Journal of Lightwave Technology}, vol.~38, no.~17, pp.
  4683--4690, 2020.

\bibitem{tang2020low}
X.~Tang, Y.-W. Chen, R.~Zhang, S.~Yao, Q.~Zhou, S.~Shen, M.~Guo, Y.~Qiao, and
  G.-K. Chang, ``{Low-complexity equalizer with a hybrid decision scheme for 50
  Gb/s/$\lambda$ PAM4-PON using a low-cost 10 G receiver},'' \emph{Optics
  Letters}, vol.~45, no.~22, pp. 6278--6281, 2020.

\bibitem{Zhou:21}
J.~Zhou, H.~Wang, Y.~Feng, W.~Liu, S.~Gao, C.~Yu, and Z.~Li, ``{Processing for
  dispersive intensity-modulation and direct-detection fiber-optic
  communications},'' \emph{Optics Letters}, vol.~46, no.~1, pp. 138--141, Jan
  2021.

\bibitem{zhou2016transmission}
J.~Zhou, L.~Zhang, T.~Zuo, Q.~Zhang, S.~Zhang, E.~Zhou, and G.~N. Liu,
  ``{Transmission of 100-Gb/s DSB-DMT over 80-km SMF Using 10-G class TTA and
  Direct-Detection},'' in \emph{ECOC 2016; 42nd European Conference on Optical
  Communication}.\hskip 1em plus 0.5em minus 0.4em\relax VDE, 2016, pp. 1--3.

\bibitem{ingle2016digital}
V.~K. Ingle and J.~G. Proakis, \emph{Digital signal processing using matlab: a
  problem solving companion}.\hskip 1em plus 0.5em minus 0.4em\relax Cengage
  Learning, 2016.

\bibitem{song2003reduced}
H.~Song and J.~Cruz, ``{Reduced-complexity decoding of Q-ary LDPC codes for
  magnetic recording},'' \emph{IEEE Transactions on Magnetics}, vol.~39, no.~2,
  pp. 1081--1087, 2003.

\bibitem{xie2009improving}
N.~Xie, T.~Zhang, and E.~F. Haratsch, ``{Improving burst error tolerance of
  LDPC-centric coding systems in read channel},'' \emph{IEEE Transactions on
  Magnetics}, vol.~46, no.~3, pp. 933--941, 2009.

\bibitem{palicot2000weighted}
J.~Palicot, ``{A weighted decision feedback equalizer with limited error
  propagation},'' in \emph{IEEE International Conference on
  Communications}.\hskip 1em plus 0.5em minus 0.4em\relax IEEE, 2000, pp.
  377--381.

\bibitem{palicot2008performance}
J.~Palicot and A.~Goupil, ``{Performance analysis of the weighted decision
  feedback equalizer},'' \emph{Signal Processing}, vol.~88, no.~2, pp.
  284--295, 2008.

\bibitem{zhou2019adaptive}
J.~Zhou, H.~Wang, J.~Wei, L.~Liu, X.~Huang, S.~Gao, W.~Liu, J.~Li, C.~Yu, and
  Z.~Li, ``{Adaptive moment estimation for polynomial nonlinear equalizer in
  PAM8-based optical interconnects},'' \emph{Optics Express}, vol.~27, no.~22,
  pp. 32\,210--32\,216, 2019.

\bibitem{bergano1993margin}
N.~S. Bergano, F.~Kerfoot, and C.~Davidsion, ``{Margin measurements in optical
  amplifier system},'' \emph{IEEE Photonics Technology Letters}, vol.~5, no.~3,
  pp. 304--306, 1993.

\end{thebibliography}
\end{document}